\documentclass[a4paper,final,conference]{IEEEtran}
\usepackage{mathpple}
\usepackage{times}

\usepackage[top=0.6in, bottom=0.7in, left=0.6in, right=0.6in]{geometry}

\usepackage{amsmath}  
\usepackage{amssymb}  
\usepackage{mathrsfs} 
\usepackage{kantlipsum,cuted}
\usepackage{multirow}
\usepackage{tabu}

\usepackage{lipsum}

\usepackage{theorem}  
\usepackage{cite}     
\usepackage{comment}  

\usepackage{upref}
\usepackage{amsfonts}

\usepackage{verbatim}

\usepackage[dvipsnames,usenames]{color}
\usepackage{enumerate}
\usepackage{multicol}
\usepackage{blkarray, bigstrut}
\usepackage{graphicx}
\usepackage{subfigure}

\usepackage{latexsym}

\usepackage[draft,ulem=normalem]{changes}
\usepackage[all]{xy}

\usepackage{algorithmicx}
\usepackage{algorithm}
\usepackage[noend]{algpseudocode}
\algrenewcommand\alglinenumber[1]{\scriptsize #1:}
\algrenewcommand\algorithmicindent{1em}%

\newcommand{\RN}[1]{%
	\textup{\uppercase\expandafter{\romannumeral#1}}%
}

\usepackage{wrapfig}

\makeatletter
\def\BState{\State\hskip-\ALG@thistlm}

\makeatother

\newcommand{\be}[1]{\begin{equation}\label{#1}}
\newcommand{\ee}{\end{equation}}

\newcommand{\bc}{\begin{center}}
\newcommand{\ec}{\end{center}}

\newcommand{\cC}{{\cal C}}

\newcommand{\cG}{{\cal G}}

\newcommand{\cS}{{\cal S}}

\newcommand{\cU}{{\cal U}}

\newcommand{\bfc}{{\boldsymbol c}}

\renewcommand{\leq}{\leqslant}

\renewcommand{\geq}{\geqslant}

\newcommand{\F}{\mathbb{F}}
\newcommand{\Fq}{\smash{\mathbb{F}_{\!q}}}

\newcommand{\Cref}[1]{Co\-rol\-la\-ry\,\ref{#1}}

\theoremstyle{plain} \theorembodyfont{\normalfont\slshape}

\newtheorem{thm}{Theorem$\!$}
\newenvironment{theorem}{\begin{thm}\hspace*{-1ex}{\bf.}}{\end{thm}}

\newtheorem{prop}[thm]{Proposition$\!$}

\newtheorem{lem}[thm]{Lemma$\!$}

\newtheorem{cor}[thm]{Corollary$\!$}

\newtheorem{cl}[thm]{Claim$\!$}
\newenvironment{claim}{\begin{cl}\hspace*{-1ex}{\bf.}}{\end{cl}}

\newtheorem{prob}[thm]{Problem$\!$}

\newtheorem{defi}[thm]{Definition$\!$}
\newenvironment{definition}{\begin{defi}\hspace*{-1ex}{\bf.}}{\end{defi}}

\theorembodyfont{\normalfont}

\newtheorem{exam}{Example$\!$}
\newenvironment{example}{\begin{exam}\hspace*{-1ex}{\bf .}}{\end{exam}}

\newtheorem{remrk}{Remark$\!$}

\newtheorem{cons}{Construction$\!$}
\newenvironment{Construction}{\begin{cons}\hspace*{-1ex}{\bf .}}{\end{cons}}

\definecolor{Codecolor}{named}{White}  

\newcommand{\Copen}{\mbox{\{\kern-5.50pt\{}}
\newcommand{\Cclose}{\mbox{\}\kern-5.50pt\}}}
\newcommand{\Cslash}{\mbox{$\backslash\kern-6.02pt\backslash$}}

\newcommand{\gc}{\cG\textmd{-}}

\begin{document}

\title{\textbf{Codes for Erasures over Directed Graphs}\vspace{-1ex}}
\author{\IEEEauthorblockN{\textbf{Lev Yohananov}}
	\IEEEauthorblockA{Dept. of Computer Science\\
		Technion-Israel Institute of Technology\\
		Haifa 32000, Israel \\
		Email: levyohananov@campus.technion.ac.il\vspace{-4ex}}
	\and
	\IEEEauthorblockN{\textbf{Eitan Yaakobi}}
	\IEEEauthorblockA{Dept. of Computer Science\\
		Technion-Israel Institute of Technology\\
		Haifa 32000, Israel \\
		Email: yaakobi@cs.technion.ac.il\vspace{-4ex}}}
\maketitle

\begin{abstract}
In this work we continue the study of a new class of codes, called \emph{codes over graphs}. Here we consider storage systems where the information is stored on the edges of a complete directed graph with $n$ nodes. The failure model we consider is of \emph{node failures} which are erasures of all edges, both incoming and outgoing, connected to the failed node. It is said that a code over graphs is a \textit{$\rho$-node-erasure-correcting code} if it can correct the failure of any $\rho$ nodes in the graphs of the code. While the construction of such optimal codes is an easy task if the field size is ${\cal O} (n^2)$, our main goal in the paper is the construction of codes over smaller fields. In particular, our main result is the construction of optimal binary codes over graphs which correct two node failures with a prime number of nodes.

\end{abstract}\vspace{-1ex}

\section{Introduction}\vspace{-0.5ex}
\renewcommand{\baselinestretch}{0.945}\normalsize\noindent

In this paper we follow up on a recent work from~\cite{YY17} which studies the correction of node failures in graphs. The main idea of the work in~\cite{YY17} was to look at information systems in which the information is represented by a graph. Such systems include for example neural networks~\cite{Hopfield:1988:NNP:65669.104422} and associative memories~\cite{6283016} that mimic the operation of the brain in the sense of storing and processing information by the associations between the information content. A distributed storage system~\cite{NetworkCoding} is another example where every two nodes can share a link with the information that is shared between them.

The work in~\cite{YY17} studied a new class of codes, called \emph{codes over graphs}. This model assumes that there are undirected complete graphs with $n$ \emph{nodes} (\emph{vertices}) and the information is stored on the undirected edges which connect every two nodes in the graph. Here, we extend this model to \emph{directed graphs} and construct codes over directed complete graphs; thus, the information is stored on the edges connecting every two nodes in the graph, including the self loops. Under this setup, there are $n^2$ edges in the graph which store $n^2$ symbols over some alphabet $\Sigma$. A code over graphs will be a set of directed graphs. Here, a \emph{node failure} corresponds to the erasure of all edges in the out- and in-neighborhood of the graph. Thus, a code over graphs will be called a \emph{$\rho$-node erasure-correcting code} if it can correct the failure of any $\rho$ nodes in each graph in the code.

The main approach in constructing these codes follows the one from~\cite{YY17} in which we use the adjacency matrix of the graph. The adjacency matrix is a square $n\times n$ matrix where its $(i,j)$-entry corresponds to the symbol stored on the edge from node $i$ to node $j$. Then, a failure of the $i$-th node in the graph translates to the erasure of the $i$-th row and the $i$-th column in the adjacency matrix of the graph. While there exist numerous constructions of array codes, most of them do not provide square matrices, but more than, they do not support the special structure of rows and columns erasure, as described above. The most relevant model to ours is the one studied by Roth~\cite{DBLP:journals/tit/Roth91} for the correction of crisscross error patterns. For crisscross error patterns it is assumed that some prescribed number of rows and columns have been erased, however the numbers of erased rows and erased columns can be different and with different indices. Therefore, this class of codes can be used to construct node-erasure-correcting codes, however, as we shall see, they will not be optimal.

For any $\rho$-node-erasure-correcting code, the failure of any $\rho$ nodes translated to $2n\rho-\rho^2$ failed edges in the graph and thus the minimum number of redundancy edges of such a code is at least \vspace{-1ex}
\begin{equation*}\label{eq:lb} \vspace{-1ex}
n^2 - (n-\rho)^2 = 2n\rho-\rho^2.
\end{equation*}
If a code satisfies this bound with equality, it will be called \emph{optimal}. Constructing optimal $\rho$-node-erasure-correcting code is an easy task if there is no restriction on the field size. For example, one can use an $[n^2, (n-\rho)^2,2n\rho-\rho^2+1]$ MDS code over a field of size at least $n^2-1$. Hence, the primary focus in this paper is the construction of node-erasure-correcting codes over a small field, and in particular binary codes.

The rest of this paper is organized as follows. In Section~\ref{sec:defs}, we formally define the graph model and codes over graphs we study in this paper. In Section~\ref{sec:single}, we present two constructions of $\rho$-node-erasure-correcting codes. The first one shows how to construct optimal codes over a field of size $q\geq n-1$. The second construction is based on the codes from~\cite{DBLP:journals/tit/Roth91} for crisscorss error patterns and generates binary codes for all $\rho$, however they are not optimal. In Section~\ref{sec:double}, we present our main result in the paper of optimal double-node-erasure-correcting codes.
Due to the lack of space some proofs of the results in the paper are omitted. \vspace{-1ex}

\section{Definitions and Preliminaries}\label{sec:defs}
In this section we formally define the codes over graphs we study in the paper. We follow similar definitions from~\cite{YY17} and modify them for directed graphs. Let $G=(V_n,E)$ be a \textit{directed graph}, where $V_n=\{v_0,v_1,\ldots,v_{n-1}\}$ is its set of $n$ \textit{nodes} (\textit{vertices}) and $E\subseteq V_n\times V_n$ is its set of \textit{edges}. A \emph{labeling function} $L$ of a graph $G$ over an alphabet $\Sigma$ is an assignment to the edges in $G$ by symbols from $\Sigma$, i.e., the labeling is a function $L:E\rightarrow \Sigma$. 

In this work we will only study complete directed graphs with self loops, that is, $E=V_n\times V_n$, with a labeling function $L:V_n\times V_n\rightarrow \Sigma$ and we will use the notation $G=(V_n,L)$ for such graphs. The \emph{adjacency matrix} of a graph $G=(V_n,L)$ is an $n\times n$ matrix over $\Sigma$ denoted by $A_G=[a_{i,j}]^{n-1,n-1}_{i=0,j=0}$, where $a_{i,j}= L(v_i,v_j)$ for all $0\leq i,j\leq n-1$. For an integer $n> 0$ we denote by $[n]$ the set $\{0,1,\ldots,n-1\}$. For a prime power $q$, the finite field of size $q$ will be denoted by ${\Fq}$. A linear code of size $n$, dimension $k$, and minimum distance $d$ over a field ${\Fq}$ will be denoted by an $[n,k,d]_q$ code.

Let $\Sigma$ be a ring and $G_1$ and $G_2$ be two graphs over $\Sigma$ with the same nodes set $V_n$. The operator $"+"$ between $G_1$ and $G_2$ over $\Sigma$, is defined by $G_1 + G_2 = G_3$, where $G_3$ is the unique graph satisfying $A_{G_1} + A_{G_2} = A_{G_3}$.
Similarly, the operator $"\cdot"$ between $G_1$ and an element $\alpha \in \Sigma$, is denoted by $\alpha\cdot G_1 = G_3$, where $G_3$ is the unique graph satisfying $\alpha \cdot A_{G_1} = A_{G_3}$.\vspace{-2ex}
\begin{definition}
	A \textit{\textbf{code over graphs}} of size $M$ and length $n$ over an alphabet $\Sigma$ is a set of directed complete graphs $\mathcal{C}_{\cG}=\{G_{i} = (V_{n},L_{i})  |  i\in[M]\}$ over $\Sigma$. We denote such a  code by $\gc(n,M)_{\Sigma}$ and in case $\Sigma =\{0,1\}$, it will simply be denoted by $\gc(n,M)$. The \textit{\textbf{dimension}} of a code over graphs $\mathcal{C}_{\cG}$ is $k_\cG= \log_{|\Sigma|}M$, the \textit{\textbf{rate}} is $R_\cG = k_\cG/n^2$, and the \textit{\textbf{redundancy}} is defined by $r_\cG = n^2 - k_\cG$. A code over graphs $\mathcal{C}_{\cG}$ over a ring $\Sigma$ will be called \textit{\textbf{linear}} if for all $G_1,G_2 \in \mathcal{C}_{\cG}$ and $\alpha,\beta \in \Sigma$ it holds  that $\alpha\cdot~G_1+\beta\cdot~G_2\in\mathcal{C}_{\cG}$. We denote this family of codes by $\gc[n,k_\cG]_\Sigma$. 
	
	A linear code over graphs whose first $k$ nodes contain the $k_\cG = k^2$ unmodified information symbols on their edges, is called a \textit{\textbf{systematic code over graphs}}. All other $n^2-k^2$ edges in the graph are called \textit{\textbf{redundancy edges}}. In this case we say that there are $k$ \textit{\textbf{information nodes}}, $r=n-k$ \textit{\textbf{redundancy nodes}}, and the number of \textit{\textbf{information edges}} is $k_\cG=k^2$. The redundancy is $r_\cG = n^2 - k^2$ and the rate is $R_\cG = {k^2/n^2}$.
	We denote such a  code by $\cS\gc[n,k]_{\Sigma}$. 
\end{definition}

\vspace{-1.5ex}
Let $G=(V_n,L)$ be a graph. For $i\in[n]$, the \emph{out-neighborhood set, in-neighborhood set}, of the $i$-th node is defined to be the set \vspace{-1ex}
$$N_i^{\mathrm{out}} = \{(v_i,v_j)\ | \ j\in[n] \}, \  N_i^{\mathrm{in}} = \{(v_j,v_i)\ | \ j\in[n] \},\vspace{-1ex}$$
 respectively, and the \emph{neighborhood set} of the $i$-th node is the set  $N_i = N_i^{\mathrm{out}}\cup N_i^{\mathrm{in}}$. Note that the $i$-th out-neighborhood set, in-neighborhood set, corresponds to the $i$-th column, row, in the adjacency matrix, respectively, and the $i$-th neighborhood set is the union of the $i$-th column and the $i$-th row in the adjacency matrix. A \textit{node failure} of the $i$-th node is the event in which all the edges in the neighborhood set of the $i$-th node, i.e. $N_i$, are erased. We will also denote this set by $F_i$ and refer to it by the \emph{failure set} of the $i$-th node. For convenience, we also define the \emph{out-failure set, in-failure set} of the $i$-th node by $F^{\mathrm{out}}_i = N_i^{\mathrm{out}}, F^{\mathrm{in}}_i = N_i^{\mathrm{in}}$, respectively.

When a node failures happens, the failed node is known and it is required to complete the values of the edges in its neighborhood set. This failure model leads us to the following definition. \vspace{-2ex}
\begin{definition}
A code over graphs is called a \textit{$\rho$\textbf{-node-erasure-correcting code}} if it can correct the failure of any $\rho$ nodes in each graph in the code.
\end{definition}
\vspace{-1ex}
The minimum redundancy $r_\cG$ of any $\rho$-node-erasure-correcting code of length $n$ is\vspace{-1ex}
\begin{equation*}\label{eq:red_bound}\vspace{-1ex}
r_\cG\geq n^2 - (n-\rho)^2 =2 n \rho  -\rho^2.
\end{equation*}
A code over graphs satisfying this inequality with equality will be called \emph{optimal}. Hence for systematic code over graphs the number of redundancy nodes is at least $\rho$. For all $n$ and $\rho$, an optimal $\rho$-node-erasure-correcting code can be constructed using an $[n^2, (n-\rho)^2,2n\rho -\rho^2+1]_q$ MDS code over a field of size $q\geq n^2-1$. The main goal of this work is to construct node-erasure-correcting codes over small fields.

\section{General Constructions of Codes over Graphs}\label{sec:single}

In the previous section we saw that optimal $\rho$-node-erasure-correcting codes are easy to construct over a field of size at least $n^2-1$. In this section we will present two constructions that reduce the large field size. Namely, in Section~\ref{subsec:optimal codes}, we will show an improvement of this last result and present constructions of optimal $\rho$-node-erasure-correcting codes over a field of size at least $n-1$. In order to further reduce the field size, in Section~\ref{subsec:MRA}, we present constructions of binary codes, however these codes will not be optimal.\vspace{-1ex}

\subsection{Optimal Codes over a Field of Size $q\geq n-1$}\label{subsec:optimal codes}
Let $G=(V_n,L)$ be a graph over a field $\F_q$ and $U\subseteq V_n\times V_n$ the subset of its edges. We define $\bfc_U\in\F_q^{|U|}$ to be a vector over $\F_q$ of length $|U|$, where its entries are the labels of the edges in the set $U$ in their lexicographical order. For example, if $U=\{(v_2,v_4),(v_1,v_0),(v_3,v_6),(v_3,v_2)\}$, then $\bfc_U= \big(L(v_1,v_0),L(v_2,v_4),L(v_3,v_2),L(v_3,v_6) \big)$. We first start with the following claim on the intersections of neighborhood sets.\vspace{-2ex}

\begin{claim}\label{lemma00}
	Let $J$ be a subset of $[n]$ of size $\rho\geq 1$. Then, the following properties hold:
	\begin{enumerate}[(a)] 
		\item For all  $\ell\in[n]$, $|  N_\ell^{\mathrm{out}}  \cap (\bigcup_{ k \in J} F^{\mathrm{in}}_k)| = \rho$.\label{subeq00}
		\item For all $m \in[n]$, $|N_m^{\mathrm{in}}\cap(\bigcup_{k \in J}  F^{\mathrm{out}}_k)| = \rho$.\label{subeq02}
		\item For all  $\ell\in[n]\setminus J$, $| N_\ell^{\mathrm{out}} \cap(\bigcup_{ k \in J} F_k)| = \rho$.\label{subeq01}
	\end{enumerate}
\end{claim}\vspace{-1ex}

We are now ready to present the construction of $\rho$-node-erasure-correcting codes. Let us consider the adjacency matrix of each graph in the code in order to explain the main idea of the construction. Each of the first $n-\rho$ columns in the adjacency matrix belongs to an $[n,n-\rho,\rho+1]_q$ MDS code, where $q\geq n-1$, and all rows belong to the same code as well. This construction is formalized as follows. \vspace{-2ex}
\begin{Construction}\label{const:rho}
	Let $n$ and $\rho$ be two positive integers such that $n > \rho$. Let $\cC$ be an $[n,n-\rho,\rho+1]_q$ MDS code, for $q\geq n-1$. 
	The code $\cC_{\cG_1}$ is defined as follows,\vspace{-1ex}
	\begin{equation*}\label{eq:parity_eq1_4}
	\mathcal{C}_{\cG_1} = \left\{ G =(V_n,L) ~\middle|~
	\begin{array}{cc}
	\forall m\in[n-\rho], & \bfc_{N_i^{\mathrm{in}}}\in\cC \\
	\forall \ell\in[n], & \bfc_{N_i^{\mathrm{out}}}\in\cC
	\end{array}
	\right\}\vspace{-0.5ex}.
	\end{equation*}
\end{Construction}
\vspace{-3ex}
\begin{theorem}\label{th:multiple1_5}
For all $\rho$ and $n$ such that $\rho<n$, the code $\mathcal{C}_{\cG_1}$ is a $\gc[n,k_\cG=(n-\rho)^2]_q$ $\rho$-node-erasure-correcting code, where $q\geq n-1$. \vspace{-1ex}
\end{theorem}

Note that the code $\cC_{\cG_1}$ can also be a systematic $\cS\gc[n,n-\rho]_q$ code where its first $n-\rho$ nodes are the information nodes. In the adjacency matrix, this corresponds to having the information symbols in the upper left $(n-\rho)\times (n-\rho)$ matrix. Then, each of the first $n-\rho$ columns in encoded systematically by a systematic encoder of the MDS code $\cC$, and then the same procedure is invoked on each of the $n$ rows.\vspace{-0.5ex}

\subsection{Binary Construction of Codes over Graphs}\label{subsec:MRA}\vspace{-0.5ex}
In this section we present constructions of binary $\rho$-node-erasure-correcting codes for arbitrary $\rho$. This construction will be based upon a construction by Roth for the correction of crisscross error patterns~\cite{DBLP:journals/tit/Roth91}. 

Let $\Gamma = [\gamma_{i,j}]^{n-1,n-1}_{i=0,j=0}$ be an $n\times n$ matrix over a field $\F$. A \textit{cover} of $\Gamma$ is defined to be a pair of two sets $(S,T)$ where $S,T \subseteq [n]$ such that for all $i,j\in[n]$ if $\gamma_{i,j}\neq 0$ then either $i\in S$ or $j\in T$,~\cite{DBLP:journals/tit/Roth91}. The \textit{cover-weight} of $\Gamma$ is defined to be the minimum size of any cover $(S,T)$ of $\Gamma$, that is, \vspace{-1ex}
$$w(\Gamma) = \min_{(S,T)\textmd{ is a cover of $\Gamma$}}\{|S|+|T|\}.\vspace{-1ex}$$ 
An $[n\times n, k, d]$ linear array code $\cC$ over a field $\F$ is a $k$-dimensional linear space of $n\times n$ matrices over $\F$, where the minimum cover-weight of all nonzero matrices in $\cC$ is $d$. It was claimed in~\cite{DBLP:journals/tit/Roth91} that the code $\cC$ can correct the erasure of any $d-1$ rows or columns in the array. Furthermore, the singleton bound for such array codes states that~\cite{DBLP:journals/tit/Roth91} $k\leq n(n-d+1)$,
and here we refer to array codes which meet this bound as \emph{optimal array codes}. In~\cite{DBLP:journals/tit/Roth91}, a construction of optimal array codes $[n\times n,n(n-r),r+1]$ was given for all $r< n$. In fact, another stronger property of this code was proved in~\cite{DBLP:journals/tit/Roth91}, in which the rank of every matrix in the code is at least $r+1$.

We are now ready to present the construction of binary $\rho$-node-erasure-correcting codes. \vspace{-2.6ex}
\begin{Construction}\label{const:rho2}
Let $\cC$ be an $[n\times n,n(n-2\rho),2\rho+1]$ binary optimal array code from~\cite{DBLP:journals/tit/Roth91}, where $\rho <n/2$. The code over graphs $\cC_{\cG_2}$ is defined as follows, \vspace{-1ex}
\begin{equation*}\label{eq:parity_eq2_1}
\mathcal{C}_{\cG_2} = \left\{ G =(V_n,L) ~\middle|~
A_G\in \cC	\right\}.
\end{equation*} 
\end{Construction}\vspace{-3.5ex}
\begin{theorem}\label{th:multiple2_2}
For all $\rho<n/2$, the code $\mathcal{C}_{\cG_2}$ is a $\gc[n,k_\cG=n(n-2\rho)]$ $\rho$-node-erasure-correcting code.
\end{theorem}\vspace{-1ex}

The construction of binary optimal array codes $[n\times n,n(n-~r),r+1]$ from~\cite{DBLP:journals/tit/Roth91} has also a systematic construction, where the first $n-2\rho$ rows of each matrix store the information bits and the last $2\rho$ rows store the redundancy bits. Therefore, we can use this family of codes also for the construction of systematic $\cS\gc[n,k=n-2\rho]$ codes over graphs for $\rho~<~n/2$.

We saw in this section that optimal codes exist for a field of size at least $n-1$, while the binary construction does not provide optimal codes. Our next task is to achieve these two properties simultaneously, that is, optimal binary codes. In the next section we show how to accomplish this task for two node failures, when the number of nodes is a prime number. The general case for arbitrary number of node failures is left for future work.\vspace{-0.5ex}

\section{Double-Node-Erasure-Correcting Codes}\label{sec:double}

In this section we present a construction of optimal binary double-node-erasure-correcting codes. We first start by reviewing a construction from~\cite{YY17} for the correction of two node failures for undirected graphs. We then show another construction of such codes for undirected graphs, and lastly we show how to combine between these two constructions in order to generate a code correcting two node failures for directed graphs. 

We refer to a graph with only undirected edges as an \textit{undirected graph}. Here, we consider only complete undirected graphs and denote them by $G_{\cU} = (V_n, L_\cU)$, where $V_n=\{v_0,v_1,\ldots,v_{n-1}\}$ is the set of $n$ nodes, and there exists an undirected edge between every two nodes, including self loops. As for the directed case, $L_\cU$ is a \emph{labeling function} that assigns every edge with a symbol over some alphabet $\Sigma$. An edge between node $v_i$ to node $v_j$ is denoted by $\langle v_i,v_j\rangle$ where the order in this pair does not matter, that is, the pair $\langle v_i,v_j\rangle$ is identical to the pair $\langle v_j,v_i \rangle$.

An undirected graph $G_\cU$ can be represented by its \emph{lower-triangle-adjacency matrix} of order $n\times n$, that is, $A_{G_\cU}~=~[a_{i,j}]^{n-1,n-1}_{i=0,j=0}$ such that $a_{i,j}=L_\cU(\langle v_i, v_j \rangle)$ if $i\geq j$ and otherwise $a_{i,j}=0$. It can also be represented by an \emph{upper-triangle-adjacency matrix} by taking the transpose of $A_{G_\cU}$.

A \textit{code over undirected graphs} of size $M$ and length $n$ over an alphabet $\Sigma$ is a set of undirected graphs $\mathcal{C}_{\cU}=\{G_{\cU_{i}} = (V_{n},L_{\cU_{i}})  |  i\in[M]\}$ over $\Sigma$.  
A \emph{linear code over undirected graphs} is defined in a similar way to the directed case and a linear code over undirected graphs whose first $k$ nodes contain the $\binom{k+1}{2}$ unmodified information symbols on their edges, is called a \textit{systematic code over undirected graphs} and will be denoted by $\cU\gc[n,k]_{\Sigma}$.

The neighborhood set of the $i$-th node in a graph $G_\cU=(V_n,L_\cU)$ is the set $N_i = \{\langle v_i, v_j \rangle | j \in [n]\}$. A \textit{node failure} of the $i$-th node is the event in which all the edges in the neighborhood set of the $i$-th node are erased and we will denote this set by $\widehat{F}_i = N_i$ to indicate a failure set.
A code over undirected graphs is called an \textit{undirected $\rho$-node-erasure-correcting code} if it can correct the failure of any $\rho$ nodes in each graph in the code.

Let $n\geq 5$ be a prime number and $G_\cU = (V_n,L_\cU)$ be an undirected graph with $n$ vertices. We use the notation $\langle a\rangle_n$ to denote the value of $(a \bmod n)$. Let us define for $h\in [n-1]$  \vspace{-2ex} 
\begin{equation*}
S_{h} = \begin{cases} 
\big\{\langle v_h,v_\ell \rangle ~|~\ell\in [n-1] \big\} &,  h \in [n-2], \\
\big\{\langle v_\ell,v_\ell \rangle ~\hspace{0.6ex}|~ \ell \in[n-1] \big\} &, h=n-2,\\
\end{cases}\vspace{-1ex}
\end{equation*}
and for $m\in[n]$,\vspace{-1ex}
\begin{equation*}
\begin{aligned}
\hspace{-0.9ex}D_{m} \hspace{-0.3ex}=\hspace{-0.2ex} & \big\{\hspace{-0.3ex} \langle v_k,\hspace{-0.3ex}v_{\ell}\rangle | k,\ell\hspace{-0.3ex}\in\hspace{-0.3ex} [n]\hspace{-0.5ex}\setminus\hspace{-0.8ex}\{n\hspace{-0.3ex}-\hspace{-0.3ex}2\},\hspace{-0.3ex}\langle k\hspace{-0.3ex}+\hspace{-0.3ex}\ell\rangle_n\hspace{-0.6ex}=\hspace{-0.3ex}m \big\} \hspace{-0.5ex}\cup\hspace{-0.5ex}  \big\{\hspace{-0.3ex}(v_{n-1},\hspace{-0.3ex}v_{n-2})\hspace{-0.3ex}\big\}\hspace{-0.3ex}.
\end{aligned}\vspace{-1ex}
\end{equation*}
The sets $S_{h}$ where $h\in[n-2]$, will be used to represent parity constraints on the neighborhood of each of the first $n-2$ nodes in the undirected graph (which correspond to ``opposite-L paths" in the lower-triangle-adjacency matrix), while the set $S_{n-2}$ will be used to impose a parity constraint on the self loops of the first $n-1$ nodes. Similarly, the sets $D_{m}$ for $m\in[n]$, will represent parity constraints on the diagonals of the lower-triangle-adjacency matrix of the graph.\vspace{-1.8ex}

\begin{example}
	The sets $S_h,D_m$ for $n=7$ are marked in Fig.~\ref{fig:graph exampleun25.1}. Entries on lines with the same color belong to the same parity constraint.
	\vspace{-3ex}
	\begin{figure}[h!]
		\hfill
		\subfigure[Neighborhood Parity Paths]{\includegraphics[width=43mm]{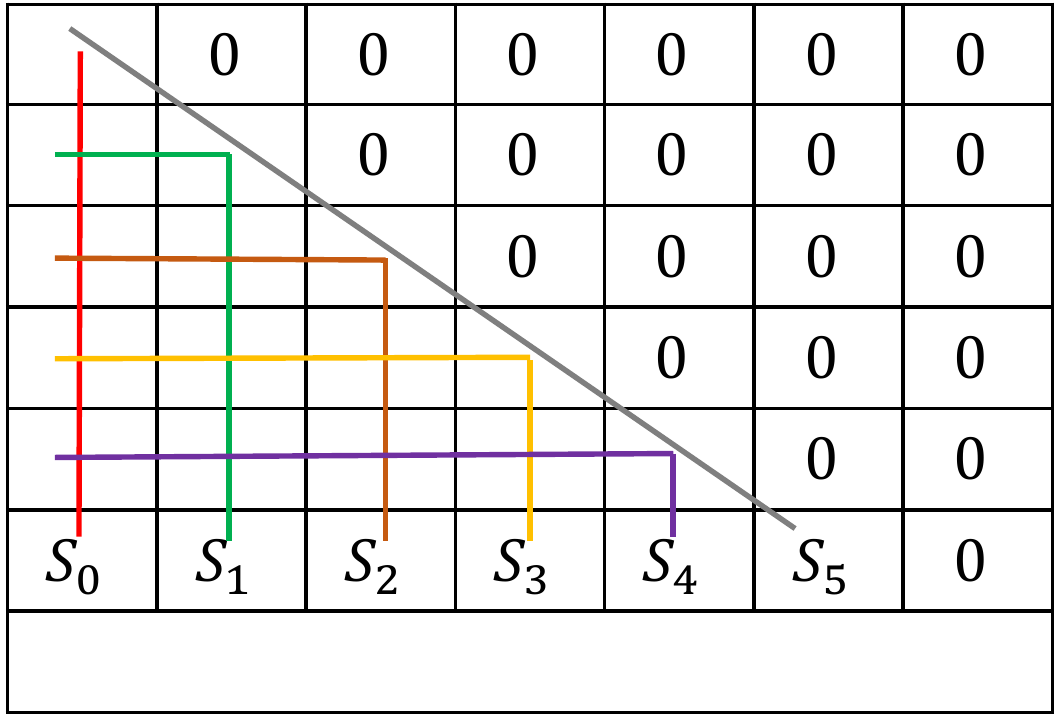}}\label{fig:graph exampleun15.1}
		\hfill
		\subfigure[Diagonal Parity Paths]{\includegraphics[width=43mm]{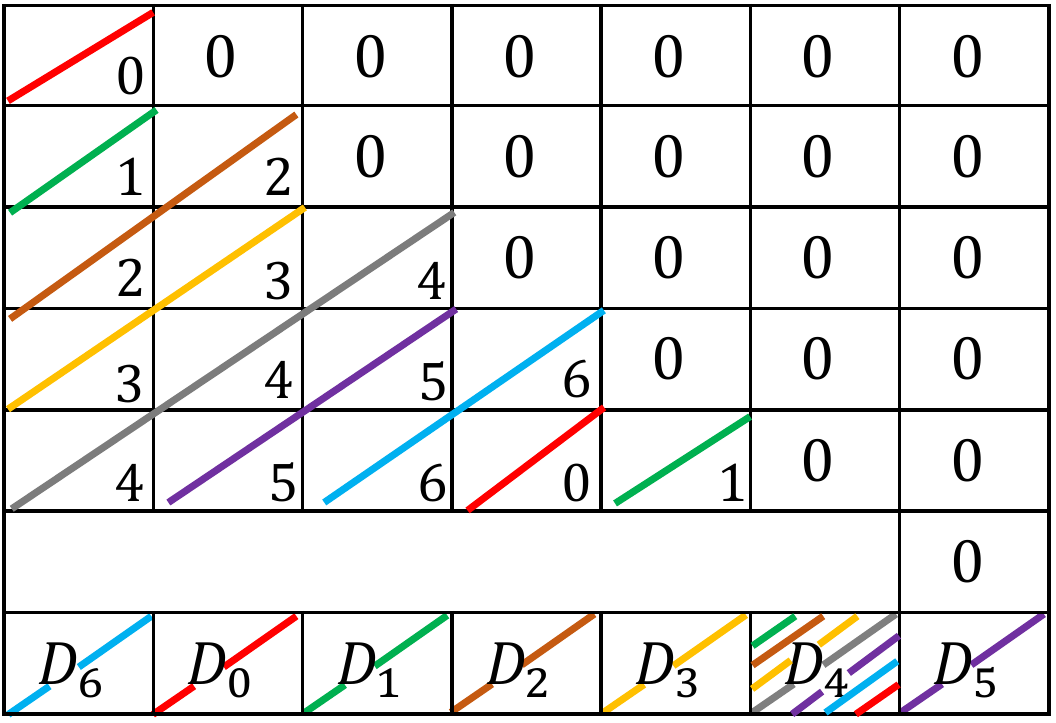}}\label{fig:graph exampleun15.2} \vspace{-3ex}
		\hfill
		\caption{The neighborhoods and diagonals sets.}\vspace{-3ex}
	\end{figure}\vspace{-0.9ex}
\end{example}

In~\cite{YY17}, we presented the following construction of systematic binary $\cU\gc[n,n-2]$ undirected double-node-erasure-correcting codes.\vspace{-2ex}
\begin{Construction}\label{const_undirected}
	For all $n\geq 5$ prime number let $\cC_{\cU_1}$ be the following code over graphs, \vspace{-1ex}
	\begin{equation*}\label{eq:parityun2_eq3}
	\hspace{-0.5ex}\mathcal{C}_{\cU_1} \hspace{-0.7ex}=\hspace{-0.7ex} \left\{ \hspace{-0.5ex}G_\cU \hspace{-0.5ex}=\hspace{-0.5ex} (V_n,L_\cU)  \middle|
	\begin{array}{cc}
	\hspace{-1ex}(a)\hspace{-0.5ex} \sum_{\langle v_i,v_j \rangle \in S_h}\hspace{-0.7ex}L_\cU(v_i,v_j)\hspace{-0.5ex}=\hspace{-0.5ex}0, h\hspace{-0.5ex}\in \hspace{-0.5ex}[n\hspace{-0.5ex}-\hspace{-0.5ex}1]  \\ 
	\hspace{-1.2ex}(b) \sum_{\langle v_i,v_j \rangle \in D_m}L_\cU(v_i,v_j)\hspace{-0.5ex}=\hspace{-0.5ex}0, m\hspace{-0.5ex}\in\hspace{-0.5ex}[n]
	\end{array}\hspace{-1.5ex}
	\right\}\hspace{-0.5ex}.\vspace{-1ex}
	\end{equation*}
\end{Construction}

The correction of this construction was proved in~\cite{YY17} by explicitly showing its decoding procedure for any two failed nodes $v_i$ and $v_j$. The more challenging case in which $i,j\in [n-2]$ works as follows. For $h \in [n-1] \setminus \{i,j\}$ and $m\in [n]$ let $\overline{S}_{h}$ and  $\overline{D}_{m}$  be the sets $\overline{S}_{h} = S_{h}\setminus~(\widehat{F}_i\cup~\widehat{F}_j)$ and $\overline{D}_{m} ~=~D_{m}~\setminus~(\widehat{F}_i\cup \widehat{F}_j)$.  Denote the \textit{syndromes} $\widehat{S}_{h},\widehat{D}_{m}$ by \vspace{-1ex}
\begin{equation*}
\widehat{S}_{h}=\hspace{-3ex} \sum_{ \substack{ \langle v_k,v_{\ell} \rangle  \in\overline{S}_{h} }} \hspace{-2ex} L_\cU(v_{k},v_{\ell}),\widehat{D}_{m}=\hspace{-3ex} \sum_{ \substack{ \langle v_k,v_{\ell}\rangle  \in\overline{D}_{m} } } \hspace{-2ex} L_\cU(v_k,v_{\ell}).\vspace{-1ex}
\end{equation*}
respectively.
Let $d = \langle j-i \rangle _{n}$, $x=\langle -1- d^{-1}\rangle_n$ and $y=\langle -1+ d^{-1}\rangle_n$. The decoding procedure for this case is presented in Algorithm~\ref{un1_algorithm}.\vspace{-1.5ex}
\begin{algorithm}
	\scriptsize
	\caption{ }
	\label{un1_algorithm}\vspace{-2ex}
	\begin{multicols}{2}
		\begin{algorithmic}[1]
			\State $b_{prev} \leftarrow 0$
			\For{$t=0,1,\ldots, x$}
			\State\hspace{-1.5ex}$s_{1} \leftarrow \langle -d(t+1)-2 \rangle _{n}$
			\State\hspace{-1.5ex}$s_{2} \leftarrow \langle s_{1} + j \rangle _{n}$
			\hspace{-1.5ex}\If{$(s_1\notin \{i,j,n-1\})$}
			\State \hspace{-1.5ex}$L_\cU(v_{s_1},v_j) \leftarrow  \widehat{D}_{s_{2}} + b_{prev}$
			\State \hspace{-1.5ex}$L_\cU(v_{s_1},v_i)\gets\widehat{S}_{s_{1}} \hspace{-0.5ex}+\hspace{-0.5ex} L_\cU(v_{s_1},v_j)$
			\State\hspace{-1.5ex}$b_{prev} \leftarrow L_\cU(v_{s_1},v_i)$ 
			\EndIf
			\hspace{-1.5ex}\If{$(s_1 = j)$}
			\State\hspace{-1.5ex}$L_\cU(v_{s_1},v_j) \leftarrow  \widehat{D}_{s_{2}} + b_{prev}$
			\State \hspace{-1.5ex}$L_\cU(v_i,v_i) \hspace{-0.5ex} \leftarrow \hspace{-0.5ex} \widehat{S}_{n-2} \hspace{-0.5ex}+\hspace{-0.5ex} L_\cU(v_{s_1},v_j)$
			\State\hspace{-1.5ex}$b_{prev} \leftarrow  L_\cU(v_i,v_i)$
			\EndIf
			\hspace{-1.5ex}\If{$s_1 = n-1$}	
			\State\hspace{-1.5ex}$L_\cU(v_{s_1},v_j) \leftarrow  \widehat{D}_{s_{2}} + b_{prev}$
			\EndIf
			\EndFor
			\State $b_{prev} \leftarrow 0$
			\For{$t=0,1,\ldots,y$}
			\State\hspace{-1.5ex}$s_{1} \leftarrow \langle d(t+1)-2 \rangle _{n}$
			\State\hspace{-1.5ex}$s_{2} \leftarrow \langle s_{1} + i \rangle _{n}$
			\hspace{-1.5ex}\If{$(s_1\notin \{i,j,n-1\})$}
			\State \hspace{-1.5ex}$L_\cU(v_{s_1},v_i) \leftarrow  \widehat{D}_{s_{2}} + b_{prev}$
			\State \hspace{-1.5ex}$L_\cU(v_{s_1},v_j) \leftarrow \widehat{S}_{s_{1}} \hspace{-0.5ex}+\hspace{-0.5ex} L_\cU(v_{s_1},v_i)$
			\State\hspace{-1.5ex}$b_{prev} \leftarrow L_\cU(v_{s_1},v_j)$ 
			\EndIf
			\If{$(s_1 = i)$}
			\State\hspace{-1.5ex}$L_\cU(v_{s_1},v_i) \leftarrow  \widehat{D}_{s_{2}} + b_{prev}$
			\State\hspace{-1.5ex}$L_\cU(v_j,v_j) \hspace{-0.5ex} \leftarrow \hspace{-0.5ex} \widehat{S}_{n-2} \hspace{-0.5ex}+\hspace{-0.5ex} L_\cU(v_{s_1},v_i)$
			\State\hspace{-1.5ex}$b_{prev} \leftarrow L_\cU(v_j,v_j)$
			\EndIf
			\If{$s_1 = n-1$}	
			\State $L_\cU(v_{s_1},v_i) \leftarrow  \widehat{D}_{s_{2}} + b_{prev}$
			\EndIf
			\EndFor
		\end{algorithmic}
	\end{multicols}\vspace{-3ex}
\end{algorithm}\vspace{-1.5ex}

Next we present another construction of systematic binary $\cU\gc[n,n-2]$ undirected double-node-erasure-correcting codes which is very similar to the codes from Construction~\ref{const_undirected}. Here, we present this construction by its constraints on its upper-triangle-adjacency matrix representation of the graphs.
For $h\in [n-1]$ denote, 
\begin{equation*}\vspace{-1ex}
S'_{h} = \begin{cases} 
\big\{\langle v_h,v_\ell \rangle ~|~\ell\in [n]\setminus\{n-2\} \big\} &,  h \in [n-2], \\
\big\{\langle v_\ell,v_\ell \rangle ~\hspace{0.6ex}|~ \ell \in[n]\setminus\{n-2\} \big\} &, h=n-2, \\
\end{cases}
\end{equation*}
and for $m\in[n]$,
\begin{equation*}
\begin{aligned}
\hspace{-0.9ex}D'_{m} \hspace{-0.3ex}=\hspace{-0.2ex} & \big\{\hspace{-0.3ex} \langle v_k,\hspace{-0.3ex}v_{\ell})\rangle | k,\ell\hspace{-0.3ex}\in\hspace{-0.3ex} [n-1]\},\hspace{-0.3ex}\langle k\hspace{-0.3ex}+\hspace{-0.3ex}\ell\rangle_n\hspace{-0.6ex}=\hspace{-0.3ex}m \big\} \hspace{-0.3ex}\cup\hspace{-0.3ex}  \big\{\hspace{-0.3ex}(v_{n-2},\hspace{-0.3ex}v_{n-1})\hspace{-0.3ex}\big\}.
\end{aligned}
\end{equation*}
As before, the sets $S'_{h}$ for $h\in[n-1]$ and $D'_{m}$ for $m \in [n]$, will be used to represent parity constraints on the upper-triangle-adjacency matrix.\vspace{-2ex}
\vspace{-0.5ex}
\begin{example}\label{example_4}
	The sets $S'_h,D'_m$ for $n=7$ are marked in Fig.~\ref{fig:graph exampleun25.1}. Entries on lines with the same color belong to the same parity constraint.
	\vspace{-3ex}
	\begin{figure}[h!]
		\hfill
		\subfigure[Neighborhood Parity Paths]{\includegraphics[width=43mm]{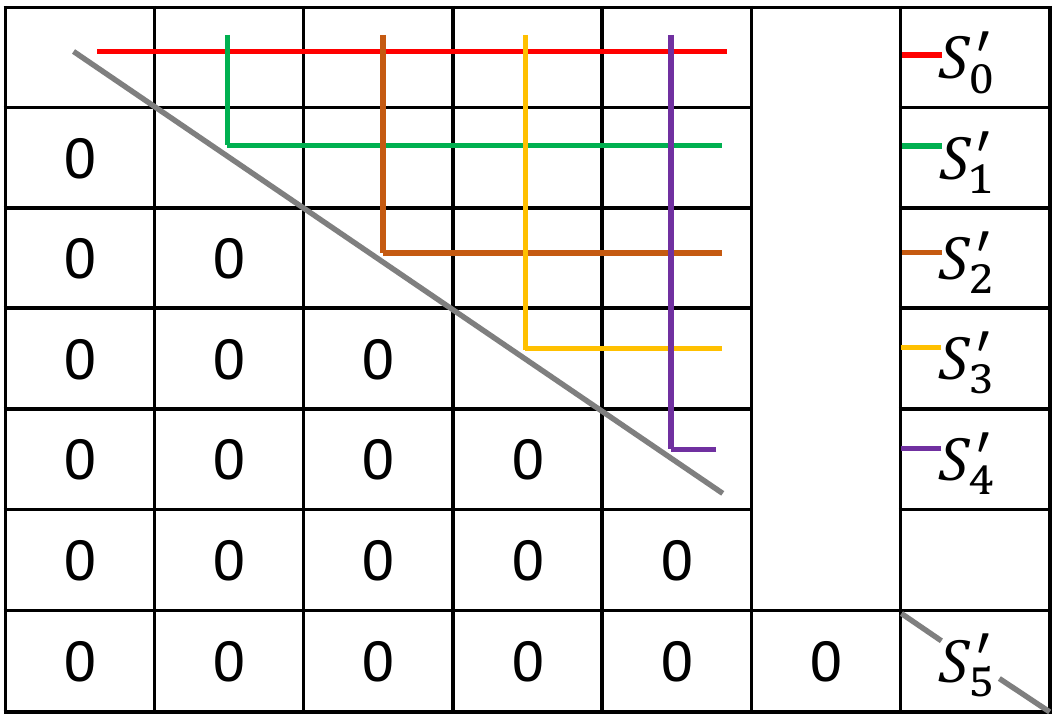}}\label{fig:graph exampleun25.1}
		\hfill
		\subfigure[Diagonal Parity Paths]{\includegraphics[width=43mm]{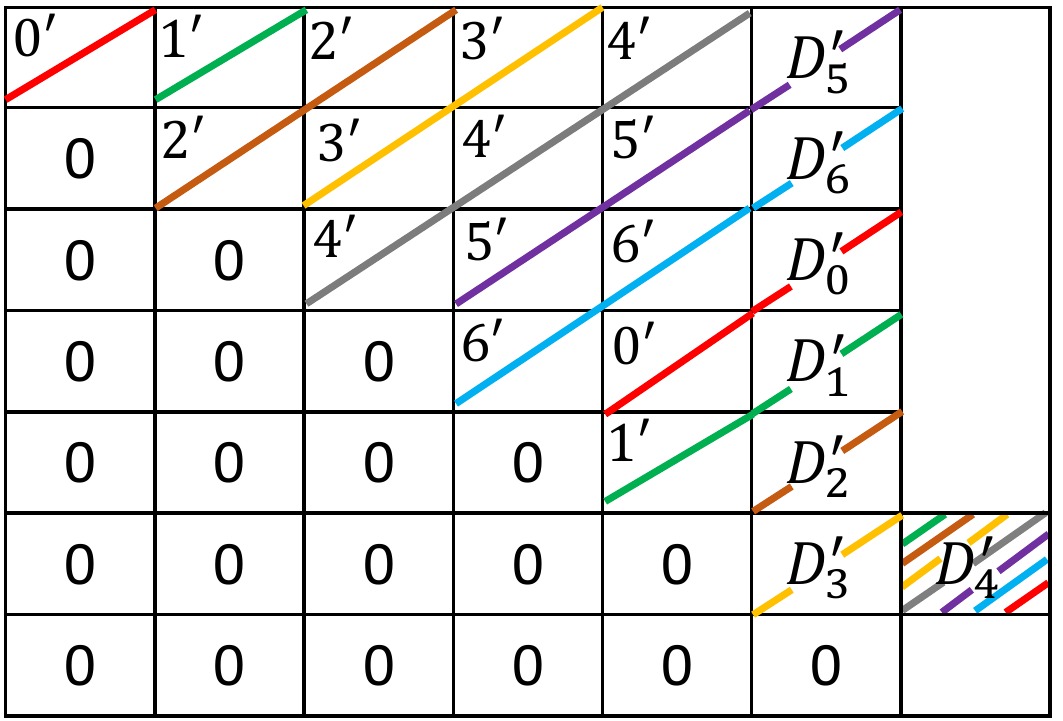}}\label{fig:graph exampleun25.2} \vspace{-3ex}
		\hfill
		\caption{The neighborhoods and diagonals sets.}\vspace{-1ex}
	\end{figure}
\end{example}\vspace{-3.5ex}

Our second construction of $\cU\gc[n,n-2]$ undirected double-node-erasure-correcting codes works as follows.
\begin{Construction}\vspace{-0.2ex}
	For all $n\geq 5$ prime number let $\cC_{\cU_2}$ be the following code over graphs, \vspace{-1ex}
	\begin{equation*}\label{eq:parityun2_eq4}
	\hspace{-0.5ex}\mathcal{C}_{\cU_2} \hspace{-0.7ex}=\hspace{-0.7ex} \left\{ \hspace{-0.5ex}G_\cU \hspace{-0.5ex}=\hspace{-0.5ex} (V_n,L_\cU)  \middle|
	\begin{array}{cc}
	\hspace{-1ex}(a)\hspace{-0.5ex} \sum_{\langle v_i,v_j \rangle \in S'_h}\hspace{-0.7ex}L_\cU(v_i,v_j)\hspace{-0.5ex}=\hspace{-0.5ex}0, h\hspace{-0.5ex}\in \hspace{-0.5ex}[n\hspace{-0.5ex}-\hspace{-0.5ex}1]  \\ 
	\hspace{-1.2ex}(b) \sum_{\langle v_i,v_j \rangle \in D'_m}L_\cU(v_i,v_j)\hspace{-0.5ex}=\hspace{-0.5ex}0, m\hspace{-0.5ex}\in\hspace{-0.5ex}[n]
	\end{array}\hspace{-1.5ex}
	\right\}\hspace{-0.5ex}.\vspace{-0.5ex}
	\end{equation*}
\end{Construction}\vspace{-1ex}

We will not prove here the correctness of the code $\cC_{\cU_2}$ since its construction is very similar to one of the code $\cC_{\cU_1}$. However, note that when constructing the code $\cC_{\cU_2}$, we switched the roles of the last two redundancy nodes such that the first node is the \textit{diagonal parity node} and the second node is the \textit{single parity node}. 
However, we still present here a decoding algorithm of this code for the more challenging case when the failed nodes are $v_i,v_j$ and $i,j\in [n-2]$. Its correctness is similar to the one of Algorithm~\ref{un1_algorithm} as done in~\cite{YY17}. For $h \in [n-1] \setminus \{i,j\}$ and $m\in [n]$ let $\overline{S}_{h}$ and $\overline{D}_{m}$  be the sets $\overline{S}_{h} = S'_{h}\setminus~(\widehat{F}_i\cup~\widehat{F}_j)$ and $\overline{D}_{m} = D'_{m}\setminus (\widehat{F}_i\cup \widehat{F}_j)$. Denote the syndromes $\widehat{S}'_{h},\widehat{D}'_{m}$ by \vspace{-1ex}
\begin{equation*}
\widehat{S}'_{h}=\hspace{-3ex} \sum_{ \substack{ \langle v_k,v_{\ell} \rangle  \in\overline{S}_{h} }} \hspace{-2ex} L_\cU(v_{k},v_{\ell}),\widehat{D}'_{m}=\hspace{-3ex} \sum_{ \substack{ \langle v_k,v_{\ell}\rangle  \in\overline{D}_{m} } } \hspace{-2ex} L_\cU(v_k,v_{\ell}).\vspace{-1ex}
\end{equation*}
Let $x'=\langle -1+d^{-1}\rangle_n$ and $y'=\langle -1- d^{-1}\rangle_n$. The decoding procedure for this case is described in Algorithm~\ref{un2_algorithm}.

In order to construct codes over directed graphs, we will use the two codes above for undirected graphs to get a family of directed systematic binary $\cS\gc[n,n-2]$ double-node-erasure-correcting codes.\vspace{-1ex}
\begin{algorithm}
	\scriptsize
	\caption{ }
	\label{un2_algorithm}\vspace{-2ex}
	\begin{multicols}{2}
		\begin{algorithmic}[1]
			\State $b_{prev} \leftarrow 0$
			\For{$t=0,1,\ldots, x'$}
			\State\hspace{-1.5ex}$s_{1} \leftarrow \langle -d(t+1)-1 \rangle _{n}$
			\State\hspace{-1.5ex}$s_{2} \leftarrow \langle s_{1} + j \rangle _{n}$
			\hspace{-1.5ex}\If{$(s_1\notin \{i,j,n-2\})$}
			\State \hspace{-1.5ex}$L_\cU(v_{s_1},v_j) \leftarrow  \widehat{D}'_{s_{2}} + b_{prev}$
			\State \hspace{-1.5ex}$L_\cU(v_{s_1},v_i)\gets\widehat{S}'_{s_{1}} \hspace{-0.5ex}+\hspace{-0.5ex} L(v_{s_1},v_j)$
			\State\hspace{-1.5ex}$b_{prev} \leftarrow L_\cU(v_{s_1},v_i)$ 
			\EndIf
			\hspace{-1.5ex}\If{$(s_1 = j)$}
			\State\hspace{-1.5ex}$L_\cU(v_{s_1},v_j) \leftarrow  \widehat{D}'_{s_{2}} + b_{prev}$
			\State \hspace{-1.5ex}$L_\cU(v_i,v_i) \hspace{-0.5ex} \leftarrow \hspace{-0.5ex} \widehat{S}'_{n-2} \hspace{-0.5ex}+\hspace{-0.5ex} L_\cU(v_{s_1},v_j)$
			\State\hspace{-1.5ex}$b_{prev} \leftarrow  L_\cU(v_i,v_i)$
			\EndIf
			\hspace{-1.5ex}\If{$s_1 = n-2$}	
			\State\hspace{-1.5ex}$L_\cU(v_{s_1},v_j) \leftarrow  \widehat{D}'_{s_{2}} + b_{prev}$
			\EndIf
			\EndFor
			\State $b_{prev} \leftarrow 0$
			\For{$t=0,1,\ldots,y'$}
			\State\hspace{-1.5ex}$s_{1} \leftarrow \langle d(t+1)-1 \rangle _{n}$
			\State\hspace{-1.5ex}$s_{2} \leftarrow \langle s_{1} + i \rangle _{n}$
			\hspace{-1.5ex}\If{$(s_1\notin \{i,j,n-2\})$}
			\State \hspace{-1.5ex}$L_\cU(v_{s_1},v_i) \leftarrow  \widehat{D}'_{s_{2}} + b_{prev}$
			\State \hspace{-1.5ex}$L_\cU(v_{s_1},v_j) \leftarrow \widehat{S}'_{s_{1}} \hspace{-0.5ex}+\hspace{-0.5ex} L_\cU(v_{s_1},v_i)$
			\State\hspace{-1.5ex}$b_{prev} \leftarrow L_\cU(v_{s_1},v_j)$ 
			\EndIf
			\If{$(s_1 = i)$}
			\State\hspace{-1.5ex}$L_\cU(v_{s_1},v_i) \leftarrow  \widehat{D}'_{s_{2}} + b_{prev}$
			\State\hspace{-1.5ex}$L_\cU(v_j,v_j) \hspace{-0.5ex} \leftarrow \hspace{-0.5ex} \widehat{S}'_{n-2} \hspace{-0.5ex}+\hspace{-0.5ex} L_\cU(v_{s_1},v_i)$
			\State\hspace{-1.5ex}$b_{prev} \leftarrow L_\cU(v_j,v_j)$
			\EndIf
			\If{$s_1 = n-2$}	
			\State $L_\cU(v_{s_1},v_i) \leftarrow  \widehat{D}'_{s_{2}} + b_{prev}$
			\EndIf
			\EndFor
		\end{algorithmic}
	\end{multicols}\vspace{-4ex}
\end{algorithm}\vspace{-2ex}

For $i,j\in [n]$, not necessarily distinct, let $\langle v_i,v_j\rangle^{\downarrow}$ be the edge directed from $v_{\max\{i,j\}}$ to $v_{\min\{i,j\}}$, i.e., $\langle v_i,v_j\rangle^{\downarrow}= (v_{\max\{i,j\}},v_{\min\{i,j\}})$, and similarly  $\langle v_i,v_j\rangle^{\uparrow} = (v_{\min\{i,j\}},v_{\max\{i,j\}})$ is the edge directed from $v_{\min\{i,j\}}$ to $v_{\max\{i,j\}}$. For $h\in [n-2]$ the \textit{neighborhood-edge sets} $S^\downarrow_h,S^\uparrow_h$ are defined by \vspace{-1.5ex}
$$S^\downarrow_h \hspace{-0.5ex}=\hspace{-0.5ex} \{\langle v_i, v_j \rangle^\downarrow | \langle v_i, v_j \rangle \in S_h\},S^\uparrow_h\hspace{-0.5ex} = \hspace{-0.5ex}\{\langle v_i, v_j \rangle^\uparrow | \langle v_i, v_j \rangle \in S'_h\}.\vspace{-1.5ex}$$
Furthermore, for $m\in [n]$ the \textit{diagonal-edge sets} $D^\downarrow_m,D^\uparrow_m$ are defined by \vspace{-1ex}
$$D^\downarrow_m \hspace{-0.5ex} = \hspace{-0.5ex} \{\langle v_i, v_j \rangle^\downarrow |  \langle v_i, v_j \rangle \hspace{-0.5ex}\in\hspace{-0.5ex} D_m\},D^\uparrow_m \hspace{-0.5ex} = \hspace{-0.5ex} \{\langle v_i, v_j \rangle^\uparrow |\langle v_i, v_j \rangle\hspace{-0.5ex} \in\hspace{-0.5ex} D'_m\},\vspace{-1ex}$$ 
and for $t\in [n]$ the \textit{failure-edge sets} $F^\downarrow_t,F^\uparrow_t$ are defined by \vspace{-1ex}
$$F^\downarrow_t\hspace{-0.5ex} =\hspace{-0.5ex} \{ \langle v_i, v_j \rangle^\downarrow | \langle v_i, v_j \rangle \in \widehat{F}_t \},F^\uparrow_t \hspace{-0.5ex}= \hspace{-0.5ex}\{  \langle v_i, v_j \rangle ^\uparrow|  \langle v_i, v_j \rangle \in \widehat{F}_t\}.$$\vspace{-6ex}
\begin{example}
	The sets $S^\downarrow_h,S^\downarrow_h,D^\uparrow_m,D^\uparrow_m$ for $n=7$ are marked in Fig.~\ref{fig:graph example5.1}. Entries on lines with the same color belong to the same parity constraint.
	\begin{figure}[h!]
		\centering\vspace{-2.5ex}
		\subfigure[Neighborhood Parity Paths]{\includegraphics[width=43mm]{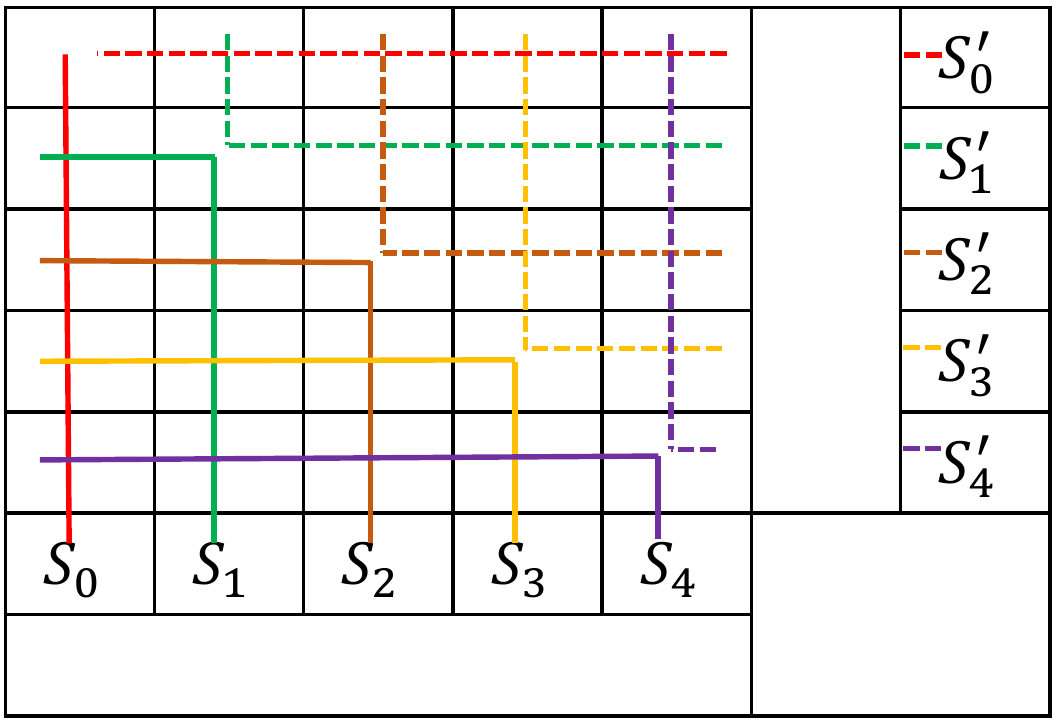}}\label{fig:graph example5.1}
		\subfigure[Diagonal Parity Paths]{\includegraphics[width=43mm]{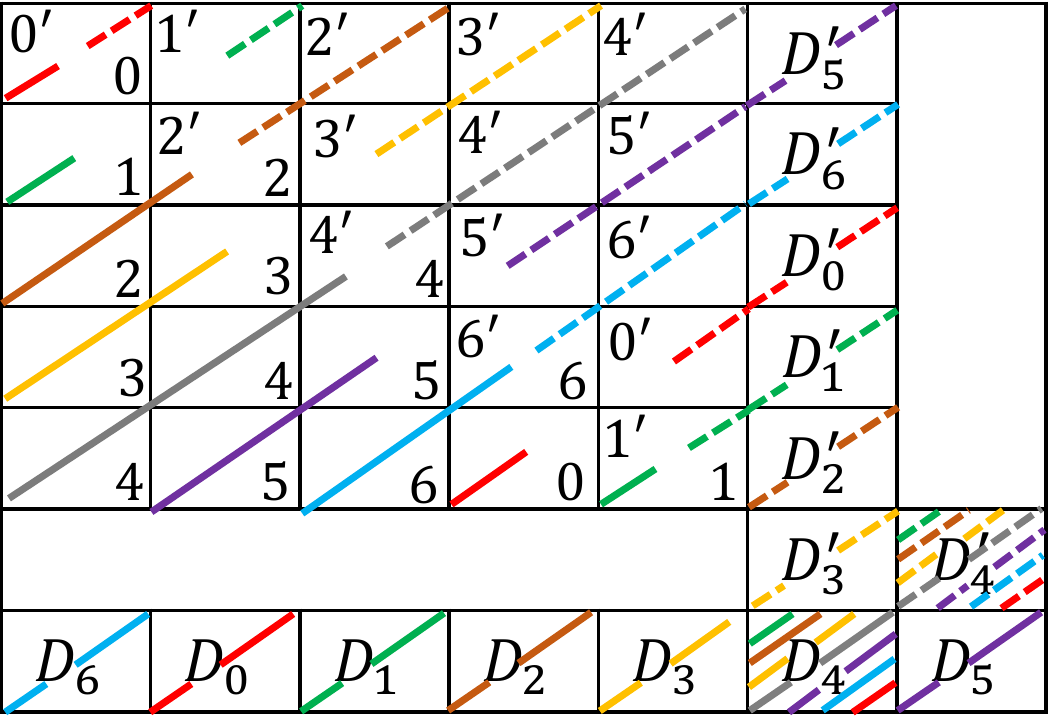}}\label{fig:graph example5.2} \vspace{-3ex}
		\caption{The neighborhoods and diagonals sets.}
	\end{figure}
\end{example}\vspace{-3.5ex}

The following claims for directed sets are very similar to the corresponding claims that were stated in~\cite{YY17}.\vspace{-2.5ex}

\begin{claim}\label{lemma0}
	For all distinct $i,j\in[n-2]$,\vspace{-1ex} $D^\downarrow_{\langle i+j\rangle_n}\cap F^\downarrow_{j} = \{( v_{j},v_{i}) \}$ and $D^\uparrow_{\langle i+j\rangle_n}\cap F^\uparrow_{j} = \{(v_{i},v_{j})\}$. \label{subeq5}
\end{claim} \vspace{-1ex}
We are now ready to present the construction of double-node-erasure-correcting codes. \vspace{-2ex}
\begin{Construction}
For all $n\geq 5$ prime number let $\cC_{\cG_4}$ be the following code.\vspace{-1ex}
\begin{equation*}\label{eq:parity_eq4}
\hspace{-1.5ex}\mathcal{C}_{\cG_4} \hspace{-0.5ex}=\hspace{-0.5ex} \left\{ \hspace{-0.5ex}G \hspace{-0.5ex}=\hspace{-0.5ex} (V_n,L)  \middle|
	\begin{array}{cc}
	\hspace{-1ex}(a)\hspace{-0.5ex} \sum_{(v_i,v_j) \in S^\downarrow_h}\hspace{-0.5ex}L(v_i,v_j)\hspace{-0.5ex}=\hspace{-0.5ex}0, h\hspace{-0.5ex}\in \hspace{-0.5ex}[n-2] \\ 
	\hspace{-1ex}(b) \sum_{(v_i,v_j)\in D^{\downarrow}_m}L(v_i,v_j)=0, m\in[n]\\ 
	\hspace{-1ex}(c)\hspace{-0.5ex} \sum_{(v_i,v_j)\in S^{\uparrow}_h}\hspace{-0.5ex}L(v_i,v_j)\hspace{-0.5ex}=\hspace{-0.5ex}0, h\hspace{-0.5ex}\in \hspace{-0.5ex}[n-2]  \\
	\hspace{-1ex}(d) \sum_{(v_i,v_j)\in D^{\uparrow}_m}L(v_i,v_j)=0, m\in[n]
	\end{array}\hspace{-1.2ex}
\right\}.
\end{equation*}

\end{Construction}\vspace{-1.8ex}
Note that in this construction we did not use the constraints that were derived from the two sets $S_{n-2}$ and $S'_{n-2}$ (i.e., the constraints on the main diagonal). Even though we do not explicitly prove it here, it is not hard to notice that this construction is systematic where the information is stored on the edges of the first $n-2$ nodes. Hence, in the next proof for the correctness of the construction we will refer to it as a systematic construction.\vspace{-2ex}

\begin{theorem}\label{th:double}
The code $\mathcal{C}_{\cG_4}$ is an optimal binary double-node-erasure-correcting code.
\end{theorem}\vspace{-1.5ex}
\begin{IEEEproof} 
Assume that nodes $i,j\in [n]$, where $i<j$ are the failed nodes. We will show the correctness of this construction by explicitly showing its decoding algorithm. We will only consider the more difficult case of $i,j\in[n-2]$.

For $h \in [n-1] \setminus \{i,j\}$ denote the sets $\overline{S}^\downarrow_{h} = S^\downarrow_{h}\setminus~(\widehat{F}^\downarrow_i\cup~\widehat{F}^\downarrow_j)$ and $\overline{S}^\uparrow_{h} = S^\uparrow_{h}\setminus~(\widehat{F}^\uparrow_i\cup~\widehat{F}^\uparrow_j)$ and for $m\in [n]$ denote the sets $\overline{D}^\downarrow_{m} = D^\downarrow_{m}\setminus (\widehat{F}^\downarrow_i\cup \widehat{F}^\downarrow_j)$ and $ \overline{D}^\uparrow_{m} =~D^\uparrow_{m}\setminus (\widehat{F}^\uparrow_i\cup \widehat{F}^\uparrow_j)$.\vspace{-0.5ex} Then, the \textit{neighborhood syndromes} $\widehat{S}^\downarrow_{h},\widehat{S}^\uparrow_{h}$ are defined by \vspace{-1ex}
\begin{equation*}
\widehat{S}^\downarrow_{h}=\hspace{-3ex} \sum_{ \substack{ \langle v_k,v_{\ell} \rangle  \in\overline{S}^\downarrow_{h} }} \hspace{-2ex} L_\cU(v_{k},v_{\ell}),\widehat{S}^\uparrow_{h}=\hspace{-3ex} \sum_{ \substack{ \langle v_k,v_{\ell}\rangle  \in\overline{S}^\uparrow_{h} } } \hspace{-2ex} L_\cU(v_k,v_{\ell}),\vspace{-1.5ex}
\end{equation*}
and the \textit{diagonal syndromes} $\widehat{D}^\downarrow_{m},\widehat{D}^\uparrow_{m}$ are defined by\vspace{-1ex}
\begin{equation*}
\widehat{D}^\downarrow_{m}=\hspace{-3ex} \sum_{ \substack{ \langle v_k,v_{\ell} \rangle  \in\overline{D}^\downarrow_{m} }} \hspace{-2ex} L_\cU(v_{k},v_{\ell}),\widehat{D}^\uparrow_{m}=\hspace{-3ex} \sum_{ \substack{ \langle v_k,v_{\ell}\rangle  \in\overline{D}^\uparrow_{m} } } \hspace{-2ex} L_\cU(v_k,v_{\ell}).\vspace{-1ex}
\end{equation*}
\vspace{-1ex}

Let $d = \langle j-i \rangle _{n}$, $x=\langle -1- d^{-1}\rangle_n$, $y=\langle -1+ d^{-1}\rangle_n$, $x'=\langle -1+d^{-1}\rangle_n$ and $y'=\langle -1- d^{-1}\rangle_n$. The decoding procedure for the code $\mathcal{C}_{\cG_4}$ in this case is described in Algorithm~\ref{directed_algorithm}.

This algorithm consists of four loops marked as Loop $\RN{1}, \RN{2}, \RN{3}$, and $\RN{4}$.
\begin{algorithm*}[t!]
	\algrenewcommand\alglinenumber[1]{\scriptsize #1:}
	\scriptsize
	\caption{ }
	\label{directed_algorithm}
	\textbf{Loop $\RN{1}$}$~~~~~~~~~~~~~~~~~~~~~~~~~~~~~~~~~~~~~~~~~~~~$\textbf{Loop $\RN{2}$}$~~~~~~~~~~~~~~~~~~~~~~~~~~~~~~~~~~~~~~~~~~~$\textbf{Loop $\RN{3}$}$~~~~~~~~~~~~~~~~~~~~~~~~~~~~~~~~~~~~~~~~~~~$\textbf{Loop $\RN{4}$}\vspace{-4ex}		
	\begin{multicols}{4}
		\begin{algorithmic}[1]
			\State $b_{prev} \leftarrow 0$
			\For{$t=0,1,\ldots, x$}
			\State\hspace{-1.5ex}$s_{1} \leftarrow \langle -d(t+1)-2 \rangle _{n}$
			\State\hspace{-1.5ex}$s_{2} \leftarrow \langle s_{1} + j \rangle _{n}$
			\hspace{-1.5ex}\If{$(s_1\notin \{i,j,n-1\})$}
			\State \hspace{-1.5ex}$L(\langle v_{s_1},v_j \rangle^\downarrow) \leftarrow  \widehat{D}^\downarrow_{s_{2}} + b_{prev}$
			\State \hspace{-1.5ex}$L(\langle v_{s_1},v_i \rangle^\downarrow) \hspace{-0.8ex}\gets \hspace{-0.8ex} \widehat{S}^\downarrow_{s_{1}} \hspace{-0.5ex}+\hspace{-0.5ex} L(\langle v_{s_1},v_j \rangle^\downarrow)$
			\State\hspace{-1.5ex}$b_{prev} \leftarrow L(\langle v_{s_1},v_i \rangle^\downarrow)$ 
			\EndIf
			\hspace{-1.5ex}\If{$(s_1 = j)$}
			\State\hspace{-1.5ex}$L(v_j,v_j) \leftarrow  \widehat{D}^\downarrow_{s_{2}} + b_{prev}$
			\State\hspace{-1.5ex}Wait until $(v_i,v_i)$ is corrected.
			\State\hspace{-1.5ex}$b_{prev} \leftarrow  L(v_i,v_i)$
			\EndIf
			\hspace{-1.5ex}\If{$s_1 = n-1$}	
			\State\hspace{-1.5ex}$L(v_{n-2},v_j) \leftarrow  \widehat{D}^\downarrow_{s_{2}} + b_{prev}$
			\EndIf
			\EndFor
			
			\State $b_{prev} \leftarrow 0$
			\For{$t=0,1,\ldots,y$}
			\State\hspace{-1.5ex}$s_{1} \leftarrow \langle d(t+1)-2 \rangle _{n}$
			\State\hspace{-1.5ex}$s_{2} \leftarrow \langle s_{1} + i \rangle _{n}$
			\hspace{-1.5ex}\If{$(s_1\notin \{i,j,n-1\})$}
			\State \hspace{-1.5ex}$L(\langle v_{s_1},v_i \rangle^\downarrow) \leftarrow  \widehat{D}^\downarrow_{s_{2}} + b_{prev}$
			\State \hspace{-1.5ex}$L(\langle v_{s_1},v_j \rangle^\downarrow) \hspace{-0.8ex} \leftarrow \hspace{-0.8ex} \widehat{S}^\downarrow_{s_{1}} \hspace{-0.5ex}+\hspace{-0.5ex} L(\langle v_{s_1},v_i\rangle^\downarrow)$
			\State\hspace{-1.5ex}$b_{prev} \leftarrow L(\langle v_{s_1},v_j \rangle^\downarrow)$ 
			\EndIf
			\If{$(s_1 = i)$}
			\State\hspace{-1.5ex}$L(v_i,v_i) \leftarrow  \widehat{D}^\downarrow_{s_{2}} + b_{prev}$
			\State\hspace{-1.5ex}Wait until $(v_j,v_j)$ is corrected.
			\State\hspace{-1.5ex}$b_{prev} \leftarrow L(v_j,v_j)$
			\EndIf
			\If{$s_1 = n-1$}	
			\State $L(v_{n-1},v_i) \leftarrow  \widehat{D}^\downarrow_{s_{2}} + b_{prev}$
			\EndIf
			\EndFor
			
			\State $b_{prev} \leftarrow 0$
			\For{$t=0,1,\ldots, x'$}
			\State\hspace{-1.5ex}$s_{1} \leftarrow \langle -d(t+1)-1 \rangle _{n}$
			\State\hspace{-1.5ex}$s_{2} \leftarrow \langle s_{1} + j \rangle _{n}$
			\hspace{-1.5ex}\If{$(s_1\notin \{i,j,n-2\})$}
			\State \hspace{-1.5ex}$L(\langle v_{s_1},v_j \rangle^\uparrow) \leftarrow  \widehat{D}^\uparrow_{s_{2}} + b_{prev}$
			\State \hspace{-1.5ex}$L(\langle v_{s_1},v_i \rangle^\uparrow)\hspace{-0.8ex} \gets \hspace{-0.8ex} \widehat{S}^\uparrow_{s_{1}} \hspace{-0.5ex}+\hspace{-0.5ex} L(\langle v_{s_1},v_j\rangle^\uparrow)$
			\State\hspace{-1.5ex}$b_{prev} \leftarrow L(\langle v_{s_1},v_i\rangle^\uparrow)$ 
			\EndIf
			\hspace{-1.5ex}\If{$(s_1 = j)$}
			\State\hspace{-1.5ex}$L(v_j,v_j) \leftarrow  \widehat{D}^\uparrow_{s_{2}} + b_{prev}$
			\State \hspace{-1.5ex}Wait until $(v_i,v_i)$ is corrected.
			\State\hspace{-1.5ex}$b_{prev} \leftarrow  L_(v_i,v_i)$
			\EndIf
			\hspace{-1.5ex}\If{$s_1 = n-2$}	
			\State\hspace{-1.5ex}$L(v_j,v_{n-2}) \leftarrow  \widehat{D}^\uparrow_{s_{2}} + b_{prev}$
			\EndIf
			\EndFor
			
			\State $b_{prev} \leftarrow 0$
			\For{$t=0,1,\ldots,y'$}
			\State\hspace{-1.5ex}$s_{1} \leftarrow \langle d(t+1)-1 \rangle _{n}$
			\State\hspace{-1.5ex}$s_{2} \leftarrow \langle s_{1} + i \rangle _{n}$
			\hspace{-1.5ex}\If{$(s_1\notin \{i,j,n-2\})$}
			\State \hspace{-1.5ex}$L(\langle v_{s_1},v_i\rangle^\uparrow) \leftarrow  \widehat{D}^\uparrow_{s_{2}} + b_{prev}$
			\State \hspace{-1.5ex}$L(\langle v_{s_1},v_j \rangle^\uparrow ) \hspace{-0.8ex} \leftarrow \hspace{-0.8ex} \widehat{S}^\uparrow_{s_{1}} \hspace{-0.5ex}+\hspace{-0.5ex} L(\langle v_{s_1},v_i\rangle^\uparrow)$
			\State\hspace{-1.5ex}$b_{prev} \leftarrow L(\langle v_{s_1},v_j \rangle^\uparrow )$ 
			\EndIf
			\If{$(s_1 = i)$}
			\State\hspace{-1.5ex}$L(v_i,v_i) \leftarrow  \widehat{D}^\uparrow_{s_{2}} + b_{prev}$
			\State\hspace{-1.5ex}Wait until $(v_j,v_j)$ is corrected.
			\State\hspace{-1.5ex}$b_{prev} \leftarrow L(v_j,v_j)$
			\EndIf
			\If{$s_1 = n-2$}	
			\State $L(v_i,v_{n-2}) \leftarrow  \widehat{D}^\uparrow_{s_{2}} + b_{prev}$
			\EndIf
			\EndFor			
		\end{algorithmic}
	\end{multicols}\vspace{-4ex}
\end{algorithm*}\vspace{0ex}
For $Y\in\{\RN{1}, \RN{2}, \RN{3},\RN{4}\}$, denote by $s^{(t)}_{1,{Y}}$ the value of the variable $s_1$ on iteration $t$ of Loop $Y$.
These values of $s^{(t)}_{1,{Y}}$ are given by:\vspace{-2ex}
\begin{align*}
&s^{(t)}_{1,{\RN{1}}} = \langle -d(t+1)-2 \rangle _{n}, s^{(t)}_{1,{\RN{2}}} = \langle d(t+1)-2 \rangle _{n},&\\\vspace{-1ex}
&s^{(t)}_{1,{\RN{3}}} = \langle -d(t+1)-1 \rangle _{n}, s^{(t)}_{1,{\RN{4}}} = \langle d(t+1)-1 \rangle _{n}.&\vspace{-1ex}
\end{align*}\vspace{-1ex}
Next, we denote the following four sets:
\begin{align}
\nonumber & A = \{s^{(t)}_{1,{\RN{1}}}:  t \in [x+1]\}, B = \{s^{(t)}_{1,{\RN{2}}}:   t \in [y+1]\},\\\vspace{-1ex}
\nonumber & A' = \{s^{(t')}_{1,{\RN{3}}}:   t'\in [x'+1]\}, B' = \{s^{(t')}_{1,{\RN{4}}}:   t \in [y'+1]\}.\vspace{-1ex}
\end{align}\vspace{-5ex}

\begin{claim}\label{claim1}
The indices $i,j$ satisfy the following property: $i,j\in A\cap B'$ or $i,j\in A'\cap B$, but not in both.
\end{claim}\vspace{-2ex}

The decoding Algorithm~\ref{directed_algorithm} for this case combines Algorithm~\ref{un1_algorithm} and Algorithm~\ref{un2_algorithm}, where Algorithm~\ref{un1_algorithm} is used to decode the lower-triangle-adjacency and Algorithm~\ref{un2_algorithm} is used to decode the upper-triangle-adjacency matrix. However, since we did not use the constraints of the two sets $S_{n-2}$ and $S'_{n-2}$ on the main diagonal, we had to replace Step 11, 25 in Algorithm~\ref{un1_algorithm}, Algorithm~\ref{un2_algorithm} with the command \textit{wait until $(v_i,v_i)$ is corrected}, \textit{wait until $(v_j,v_j)$ is corrected}, respectively. 
According to Claim~\ref{claim1}, the indices $i,j$ satisfy $i,j\in A\cap B'$ or $i,j\in A'\cap B$ but not both. Without loss of generality, assume that $i,j\in A\cap B'$. Therefore, in this case, Loops $\RN{2}$ and $\RN{3}$ of Algorithm~\ref{directed_algorithm} will not be affected by the main diagonal constraint. This holds since the edges $(v_i,v_i)$ and $(v_j,v_j)$ are not corrected in these two loops as the conditions in Steps 23 and 37 will not hold. Hence, these two loops operate and succeed exactly as done in Algorithm~\ref{un1_algorithm} and Algorithm~\ref{un2_algorithm}. This does not hold for Loops $\RN{1}$ and $\RN{4}$. Namely, Loop $\RN{1}, \RN{4}$ operates exactly as Algorithm~\ref{un1_algorithm}, Algorithm~\ref{un2_algorithm} until Loop $\RN{1}, \RN{4}$ reaches Step 11, 53, respectively. Here we notice that according to Algorithm~\ref{un1_algorithm}, in Step 11, the algorithm was supposed to correct the edge $(v_i,v_i)$ according to the constraint on the mail diagonal. Similarly, in Step 53, the algorithm was supposed to correct the edge $(v_j,v_j)$ according to the constraint on the mail diagonal. However, since the edge $(v_i,v_i)$ is corrected in Loop $\RN{4}$ and the edge $(v_j,v_j)$ is corrected in Loop $\RN{1}$, all we need to do in Step 11 is to wait for the edge $(v_i,v_i)$ to be corrected and in the same way in Step 53 for the edge $(v_j,v_j)$ to be corrected. Then, the rest of these two loops proceed to correct the remaining edges as done in Algorithm~\ref{un1_algorithm} and Algorithm~\ref{un2_algorithm}.

Lastly, from Claim~\ref{lemma0}, $D^\downarrow_{\langle i+j\rangle_n}\cap F^\downarrow_{j} = \{( v_{j},v_{i}) \}$ and $D^\uparrow_{\langle i+j\rangle_n}\cap F^\uparrow_{j} = \{(v_{i},v_{j})\}$, so the last two information edges $(v_j,v_i)$ and $(v_i,v_j)$ are corrected by constraints $\widehat{D}^\downarrow_{i+j}$ and $\widehat{D}^\uparrow_{i+j}$, respectively. Since all of the information edges were corrected, we can correct the remaining uncorrected redundancy edges $(v_{n-2},v_i)$,$(v_{n-2},v_j)$,$(v_i,v_{n-1})$ and $(v_j,v_{n-1})$ using our encoding rules.
\end{IEEEproof}

The decoding algorithm presented in the proof of Theorem~\ref{th:double} is demonstrated in the next example.\vspace{-2.3ex}
\begin{example} 
	\begin{figure}[h!]
		\centering\vspace{-2ex}
		\subfigure[Simulation of the algorithm]{\includegraphics[width=57mm]{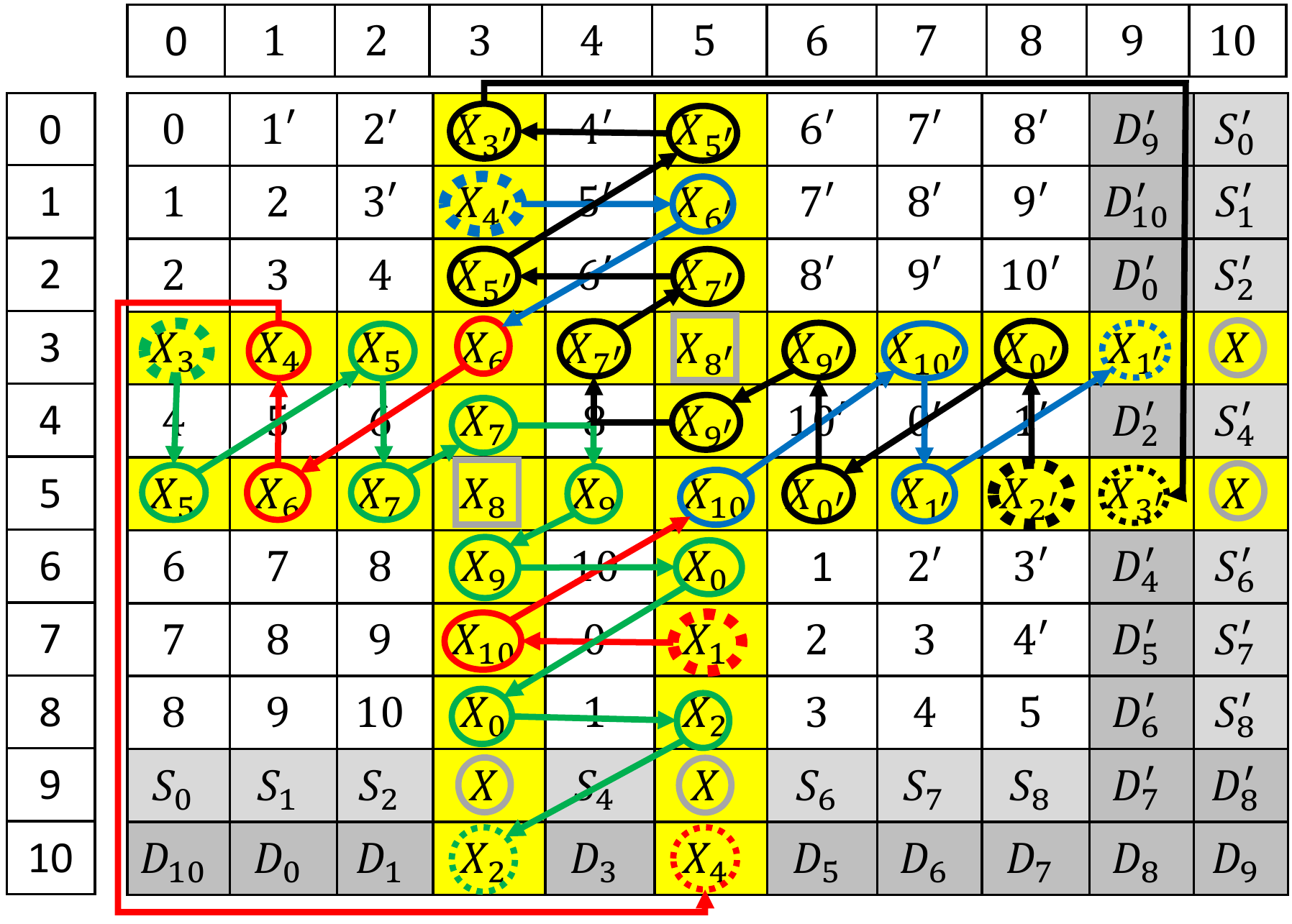}}\\\vspace{-1.5ex} 
		\subfigure[Lower tringle corrected edge order]{\includegraphics[width=36.4mm]{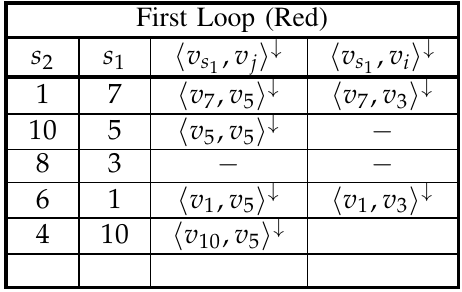}\vspace{-1ex}
		\includegraphics[width=36mm]{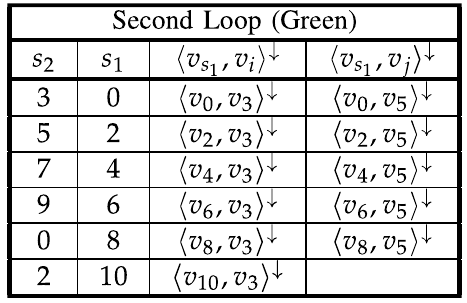}}\vspace{-1ex}
		\subfigure[Upper tringle corrected edge order]{\includegraphics[width=36mm]{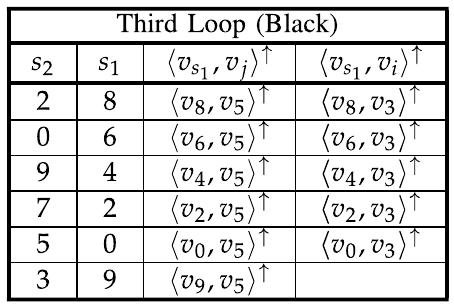}\vspace{-1ex}
		\includegraphics[width=37.4mm]{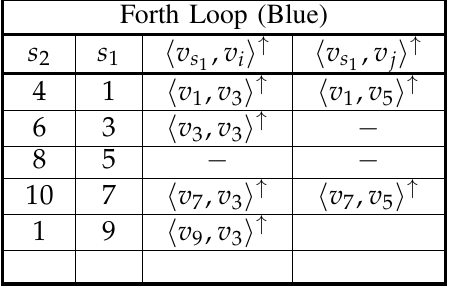}}\vspace{-1ex}
		\caption{We consider the case where $n=11$ and the failed nodes are $v_3$ and $v_5$, that is, $i=3,j=5$. Therefore $d=2$ and $x=y'=4, x'=y=5$. We use here the lower-triangle-adjacency matrix for Loop $\RN{1}$ (red) and Loop $\RN{2}$ (green) and the upper-triangle-adjacency matrix for Loop $\RN{3}$ (black) and Loop $\RN{4}$ (blue). Loop $\RN{1}$ starts with the edge $(v_7,v_5)$, and ends with the edge $(v_{10},v_5)$ and Loop $\RN{2}$ starts with the edge $(v_3,v_0)$ and ends with the edges $(v_{10},v_3)$. Similarly, Loop $\RN{3}$ starts with the edge $(v_5,v_8)$, and ends with the edge $(v_5,v_9)$ and Loop $\RN{4}$ starts with the edge $(v_1,v_3)$ and ends with the edges $(v_3,v_9)$. Loop $\RN{1}, \RN{4}$ corrects the self loop $(v_5,v_5), (v_3,v_3)$, respectively. At the end of this algorithm, $(v_5,v_3),(v_9,v_3),(v_9,v_5)$ are the uncorrected edges for the lower-triangle-adjacency matrix and $(v_3,v_5),(v_3,v_{10}),(v_5,v_{10})$ for the upper-triangle-adjacency matrix, and are marked in gray.}\vspace{-1ex}
	\end{figure}	
\end{example} \vspace{-2ex}

\bibliographystyle{abbrv}\vspace{-1ex}
\bibliography{references}

\end{document}